\documentclass[a4paper,11pt]{article}
\pdfoutput=1 

\usepackage{jcappub} 

\usepackage[T1]{fontenc} 

   \newcommand{\exclude}[1]{}
\newcommand{\be}{\begin{eqnarray}}
\newcommand{\ee}{\end{eqnarray}}

\usepackage{graphicx,amsmath,amssymb,bm}
\usepackage{psfrag}
 \usepackage{feynmp}
\usepackage{hyperref}
\usepackage{enumitem}

\title{\boldmath  511 keV Line Emission from Nearby Spherical Dwarf Galaxies} 
\author[a,1]{Kyle Lawson\note{Corresponidng author.}} 
\author[a]{Ariel  Zhitnitsky}
\affiliation[a]{Department of Physics \& Astronomy, University of British Columbia, 
Vancouver, B.C. V6T 1Z1, Canada}

 \abstract{The observed galactic 511 keV line has been interpreted in a number of papers as a possible 
signal of dark matter annihilation within the galactic bulge. If this is the case then we should 
expect a similar spectral feature associated with nearby dwarf galaxies 
which are dark matter dominated. It has recently been argued \cite{Siegert:2016ijv} 
that the absence of such a signal excludes a dark matter explanation as the  major 
source for the galactic 511 keV line.
 In the model presented here dark matter in the form of heavy 
quark nuggets produces the galactic 511 keV emission line through interactions 
with the visible matter. It is argued, however,  that this type of interaction is not subject to the 
strong dark matter annihilation constraints presented in  \cite{Siegert:2016ijv}.
}

\begin{document}
\maketitle
\flushbottom
\section{Introduction}

The origin of the galactic 511 keV positron annihilation line remains an 
outstanding problem in galactic 
astrophysics in terms of both morphology and intensity. Some details of this 
feature of the galactic spectrum and of its history can be found in the review 
\cite{Prantzos:2010wi}. Of primary interest is the low disk-to-bulge intensity ratio 
of the 511 keV emission and the apparently low injection energy of the positrons 
responsible. The absence of a clear astrophysical source combined with the largely 
spherical morphology lead to the suggestion that the primary source of the required 
population of low energy positrons could be the annihilation of light dark matter. For the 
details of this argument see the original paper \cite{Boehm:2003bt} and the many 
related references for followup papers found in the review article \cite{Prantzos:2010wi}. 
The models reviewed in \cite{Prantzos:2010wi} include a number of models in which 
the low energy positrons responsible for the 511 keV line are produced in the annihilation 
of light weakly interacting massive particles (WIMPs) and it is this class of models to 
which the analysis of \cite{Siegert:2016ijv} is most readily applicable. 

Following the detection of a stronger than expected galactic 511 keV line it
was noted \cite{Hooper:2003sh} that the dwarf satellite galaxies of the 
Milky Way offer a test of its potential dark matter origin. 
This suggestion motivated the analysis of \cite{Siegert:2016ijv} in which it was shown 
that observations from the SPI spectrometer on INTEGRAL of nearby dwarf 
galaxies contradict the basic idea that the the 
galactic 511 keV line could be related to dark matter annihilation.  In a purely J-factor 
based analysis a dark matter generated galactic 511 keV line would require similar 
emission from the dwarf galaxies at levels significantly above those observed. 
 
We consider an alternative dark matter based interpretation of the galactic 511 keV 
line consistent with both the presence of a disk component in the galactic line's 
morphology and with the non-detection of 511 keV emission from the dwarf satellite 
galaxies. This model \cite{Zhitnitsky:2002qa,Oaknin:2003uv} was originally proposed 
to offer a natural explanation 
of the observed relation between the dark and visible components of the  cosmic 
energy density $\Omega_{\rm dark} \sim \Omega_{\rm visible}$, in 
contrast most other proposals considered in this context have been specifically 
designed to produce the observed galactic 511 keV line.  
We will review the basic properties of this model 
in Section \ref{sec:QNDM}. For now it is sufficient to mention that 
in this model the dark matter consists of macroscopically large composite 
objects composed of light standard model quarks or antiquarks. These ``nuggets" of antiquarks 
are surrounded by a layer of low energy positrons which, if these objects 
serve as the dark matter, will occasionally annihilate  
with the electrons of the interstellar medium 
and contribute to the observed emission \cite{Oaknin:2004mn, Zhitnitsky:2006tu}. In this case 
the strength of the 511 keV emission is dependent on both the visible 
$n_{\rm visible}$ and dark matter $n_{\rm DM}$
densities, which is in huge contrast with conventional dark matter models in which 
the annihilation rate is determined by the dark matter density 
$n_{\rm DM}$ only. From the reliance of emission on the local visible matter 
density in this model it is  obvious that the dwarf satellite galaxies should be 
weak 511 keV sources as they are relatively poor in interstellar matter.

Following a brief review of the quark nugget dark matter model in section 
\ref{sec:QNDM} we will provide an analysis of galactic 511 keV line emission within 
the context of this model in section \ref{sec:511}. Finally, in section  \ref{sec:dwarf} we will use the 
511 keV line strength from the galactic centre to extrapolate to emission from 
nearby dwarf spherical galaxies and demonstrate that the expected emission is 
fully consistent with present constraints. Section \ref{sec:conclusion} is our conclusion.   

\section{Quark nugget dark matter}
\label{sec:QNDM}

In this section we provide a brief overview of the quark nugget 
dark matter model. Some further details on observational constraints are provided in the short review paper 
\cite{Lawson:2013bya} and the references therein. Details of a possible formation 
mechanism for the quark and antiquark nuggets is provided in 
the recent paper \cite{Liang:2016tqc}, see also \cite{Zhitnitsky:2016cir} with  a short overview  on the formation stage of the nugget's evolution   during the QCD transition in early Universe.

The idea that the dark matter may take the form of massive composite objects composed of 
standard model quarks in a novel phase goes back to stranglet models \cite{Witten:1984}. 
However the model we consider here (originally formulated in 
\cite{Zhitnitsky:2002qa,Oaknin:2003uv}) has some important distinctions from strangelet 
dark matter. First of all, the nuggets are formed of matter as well as antimatter 
as a result of separation of charges.  Secondly,  the stability of the DM nuggets is 
provided by extra pressure generated by axion domain walls, in contrast with 
strangelets for which stability is assumed to be achieved even in vacuum, at 
zero external pressure. Finally, an overall coherent baryon asymmetry in the entire 
Universe results from the strong CP violation due to the fundamental $\theta$ 
parameter in QCD which is assumed to be nonzero at the beginning of the QCD 
phase transition. 
 
Unlike conventional dark matter candidates such as WIMPs the dark matter/antimatter
nuggets are strongly interacting but macroscopically large.  
They do not contradict known observational constraints on dark matter 
or antimatter for three main reasons~\cite{Zhitnitsky:2006vt}:
\begin{itemize} 
\item They carry a huge (anti)baryon charge 
$|B| \gtrsim 10^{25}$, and so have an extremely tiny number density; 
\item The nuggets have nuclear densities, so their effective interaction
is small $\sigma/M \sim 10^{-10}$ ~cm$^2$/g,  well below the typical astrophysical
and cosmological limits which are on the order of 
$\sigma/M<1$~cm$^2$/g;
\item They have a large binding energy such that baryon charge in the
nuggets is not available to participate in big bang nucleosynthesis
(\textsc{bbn}) at $T \approx 1$~MeV. 
\end{itemize} 
To reiterate: the weakness of the visible-dark matter interaction is achieved 
in this model due to the small geometrical parameter $\sigma/M \sim B^{-1/3}$ 
rather than due to the weak coupling 
of a new fundamental field with standard model particles. In other words, this small 
effective interaction $\sim \sigma/M \sim B^{-1/3}$ replaces a conventional requirement
of sufficiently weak interactions of the visible matter with  WIMPs. 

While the observable consequences of this model are on average strongly suppressed  
by the low number density of the quark nuggets the interaction of these objects 
with the visible matter of the galaxy will necessarily produce observable 
effects. Any such consequences will be largest where the densities 
of both visible $n_{\rm visible}$ and dark matter $n_{\rm DM}$ are largest such as the 
core of the galaxy. In other words, the nuggets behave as   
conventional cold dark matter in the environment where the visible matter density is 
small, while they become interacting and radiation emitting objects (i.e. effectively 
become visible matter) when in an environment of sufficiently large density.

The rate of annihilations between visible matter and antiquark nuggets is proportional to 
the product of the local visible and DM distributions at the annihilation site. 
The observed flux for any form of annihilation driven emission thus depends on 
the same line-of-sight integral 
\be
\label{flux1}
\Phi \sim R^2\int d\Omega dl [n_{\rm visible}(l)\cdot n_{DM}(l)],
\ee
where $R\sim B^{1/3}$ is a typical size of the nuggets which determines the 
effective interaction cross section between the dark and visible matter. As 
$n_{DM}\sim B^{-1}$ the effective interaction is suppressed by a factor of $\sim B^{-1/3}$ 
as mentioned above. The form of expression \ref{flux1} should be contrasted 
with the J-factor line-of-sight integral used in \cite{Siegert:2016ijv} and many other 
analysis of dark matter related emission, 
\begin{equation}
\label{J}
J \equiv \int_{\Delta \Omega} {\rm{d}}\Omega \int \rho_{DM}^2 dl
\end{equation}
which depends strictly on the dark matter density and thus characterizes  
dark matter-dark matter interactions along a given line of sight. The 
fact that expression (\ref{flux1}) mixes the dark and visible matter distributions is 
important to the arguments which follow. The dependence on the more strongly 
clumped visible matter density make the flux predictions of this model more difficult 
despite the seemingly simple dependance on a single parameter $\langle B\rangle$. 
The estimation $\langle B\rangle\sim 10^{25}$ can be fixed 
by assuming that the annihilation of electrons from the visible matter with the positrons 
from the antiquark nuggets saturates the observed galactic 511 keV line 
\cite{Oaknin:2004mn, Zhitnitsky:2006tu}. It has also been assumed here that the 
observed dark matter density is saturated by the combination of quark and 
antiquark nuggets.   

The annihilation of galactic matter within an antiquark nugget is a complex many 
body problem which involves a range of annihilation channel each giving rise to 
emission in a different frequency band. 
Emission strengths in different bands are expressed in terms of the same 
integral (\ref{flux1}), and therefore, the relative intensities are completely determined 
by the internal structure of the nuggets which is described by conventional nuclear physics 
and basic QED.   Estimates of the nugget contribution to the diffuse galactic spectrum 
can be found in the original references 
\cite{Oaknin:2004mn, Zhitnitsky:2006tu,Forbes:2006ba, 
Lawson:2007kp,Forbes:2008uf,Forbes:2009wg,Lawson:2012zu},
where predictions of the model have been confronted with observations in specific 
frequency bands covering more than eleven orders of magnitude, from radio frequency 
with $\omega\sim 10^{-4} $ eV to $\gamma$ rays with $\omega\sim 10$ MeV.    
It has been shown that there are no contradictions with available data. Furthermore, there are 
a number of frequency bands where some excess emission is observed, but not explained 
by conventional astrophysical sources. It has been argued that the contribution of 
emission from the antiquark nuggets may explain, either wholly or in part these observed 
excesses in the galactic diffuse emission, see the original works \cite{Forbes:2006ba, 
Lawson:2007kp,Forbes:2008uf,Forbes:2009wg,Lawson:2012zu} for details.  

\section{The galactic 511 line}
\label{sec:511}
Here we present a simplified analysis of the galactic 511 keV line strength in the 
context of the quark nugget dark matter model. The annihilation of visible matter 
with the nuggets is a well understood process entirely grounded in known QED and 
QCD physics, and these processes will operate identically in a wide variety of astrophysical 
environments. In the outer regions of the electrosphere of an antiquark nugget the positrons 
carry velocities at the thermal (eV) scale and low energy galactic electrons entering 
the electrosphere will rapidly form positronium bound states \cite{Zhitnitsky:2006tu}. 
One quarter of positronium decays result in the production of a narrow 511 keV line 
while the remaining fraction contribute to a three photon continuum also observed from 
the galactic centre. 

The rate at which matter annihilates with the nuggets to produce a 511 keV photon 
pair in a given environment is approximately given by, 
\begin{equation}
\frac{dN}{dt~dV} = f_{511} \frac{\sigma}{M} \rho_{\rm DM} \langle v \rangle n_e
\end{equation}
where $f_{511}$ is the probability that a collision results in the emission of 
a 511 keV photon, $\sigma/M$ is the average cross-section to mass ratio of the 
nuggets, $\rho_{\rm DM}\simeq n_{\rm DM} M$ is the dark matter mass density, 
$\langle v \rangle$ is the 
averaged relative velocity between the nuggets and the visible matter and $n_e$ is the 
number density of electrons (both free and bound) in the interstellar medium. As we 
are interested in the relative strength of 511 keV emission from the Milky Way and 
other nearby galaxies we will define the coefficient $\kappa$ to stand in for the  
unknowns in the microscopic details of the nuggets. 
\begin{equation}
\frac{dN}{dt~dV} \equiv \kappa \rho_{\rm DM}  \langle v \rangle \rho_{\rm ISM}
\end{equation}
where $\rho_{\rm ISM}$ is the mass density of the interstellar medium\footnote{It 
should be noted that stellar matter does not contribute significantly to the production 
of 511 keV photons as photons produced within a star will be rapidly absorbed.}. 
The total annihilation rate of positrons within the galactic centre is estimated to be 
$\Gamma_{e^+} = 2\times 10^{43}$ s$^{-1}$ which allows us to fix the value of the 
coefficient $\kappa$ to be, 
\begin{equation}
\label{eq:coefficient}
\kappa = \frac{\Gamma_{e+}}{\int dV \rho_{\rm DM}  \langle v \rangle \rho_{\rm ISM}}
\end{equation}
The integral appearing in equation (\ref{eq:coefficient}) should run over the matter 
distribution of the galactic centre which is rather complicated. As a first simplified 
estimate consider the case where the dark matter has a roughly constant density 
core out to the kiloparsec scale and the collisional velocity is fixed around the typical 
galactic scale $v_g \approx 200$km/s. In this case the integral runs over only the 
visible matter distribution and we have,
\begin{equation}
\kappa = \frac{\Gamma_{e+}}{\rho_{\rm DM} v_g M_{\rm ISM}}
\end{equation}
Where $M_{\rm ISM}$ is the total mass of interstellar gas within the galactic centre. Note 
that a more strongly cusped dark matter halo would increase the value of the integral 
in the denominator of expression (\ref{eq:coefficient}) and result in a lower estimate of 
$\kappa$. Such an estimate would represent an increased suppression of 511 keV emission 
and sets an upper limit on emission per nugget and allows us to estimate 
the maximum 511 keV flux expected from nearby dwarf galaxies\footnote{A more careful 
treatment of the matter distribution of the galactic centre would allow us to better constrain 
the emission from nuggets within the Milky Way but is not necessary for present purposes 
as even this relatively high estimate of nugget emission will be shown to be 
consistent with constraints from the dwarf galaxies.}. 

In keeping with our intention to make a conservative estimate of the emission from 
nearby spherical dwarfs we will adopt a relatively low dark matter density of 
$\rho_{\rm DM} \approx 1$GeV/cm$^{-3}$ and a galactic velocity of $v_g \approx 200$km/s. 
The visible matter mass in the galactic bulge is estimated to be on the order of 
$10^{10}M_{\odot}$ \cite{McMillan:2011}. Using these estimates we arrive at a value
for the emission coefficient of, 
\begin{equation}
\label{eq:K_MW}
\kappa \approx  10^{-31} \frac{{\rm{cm}}^2}{{\rm{GeV}}^2} 
\end{equation} 

\section{Dwarf Galaxies} 
\label{sec:dwarf}
We now want to determine the 511 keV line signal associated with 
a dwarf galaxy. These objects are below the resolution of the SPI and will 
consequently appear as point sources in the INTEGRAL data.  In this case we can 
write the total flux received from a galaxy at distance $d$ as,  
\begin{equation}
\label{eq:flux_dwarf}
\frac{dN_{511}}{dt~dA} = \frac{\kappa}{4\pi d^2} \int dV~\rho_{\rm ISM} v \rho_{\rm DM}
\end{equation}
with the integration running over the volume containing the galaxy.
In order to extract flux estimates we must consider possible matter profiles 
for the dwarf galaxies\footnote{Unfortunately a model independent comparison with the 
results of \cite{Siegert:2016ijv} is not possible due to the fact that the matter integral 
appearing in expression (\ref{eq:flux_dwarf}) contains both dark and visible matter factors 
as opposed to simply the J-factor, $J\sim \rho_{\rm DM}^2$.}. As above we will attempted to 
make assumptions which 
produce conservative constraints (that is models which produce the largest 
possible flux).  To this end we will assume that the dark matter of the dwarf galaxies 
follows a conventional NFW profile \cite{Navarro:1995iw} which will 
produce stronger central intensity than alternative cored distributions, 
\begin{equation}
\label{eq:NFW}
\rho_{NFW}(r) = \frac{\rho_s}{r/r_s (1 + r/r_s)^2}
\end{equation}
for a scale radius $r_s$ and density $\rho_s$. 

The situation is more complicated for the visible matter. The dwarf galaxies are known 
to contain very little gas \cite{Grcevich:2009gt} which is the visible matter 
component of most relevance to diffuse 
511 keV emission. The stellar distribution in dwarf spheroid galaxies has been 
modelled using a Plummer profile, 
\begin{equation}
\label{eq:plummer}
\rho = \rho_0 \left[ 1 + \left( \frac{r}{r_{1/2}} \right)^2 \right]^{-5/2}
\end{equation}
where $\rho_0$ is the central density and $r_{1/2}$ is the half-light 
radius of the stellar population (see for example \cite{Irwin:1995} for a discussion 
of different stellar profiles.) 
We will assume that what ISM components are present follow a similar 
profile and mass scale though in reality the gas component could be 
much smaller. 
Alternative matter distributions have also been considered  
including exponential profiles and tidally stripped King profiles which have been 
shown to provide a good fit to star counts \cite{Irwin:1995}. It has also been suggested 
that a better fit to the Draco dwarf spheroid is obtained from a modified 
Plummer profile which falls as $\sim r^{-7/2}$ at large radii \cite{Mashchenko:2005bj}. 
However the Plummer profile is known to provide an adequate fit to a large number 
of dwarf spheroids and the relatively strong central peak of the expression (\ref{eq:plummer}) 
should produce a correspondingly strong 511 keV emission line when convolved 
with the NFW profile as will its large spacial extent relative to the truncated King profile. 

The mean velocity of the gravitationally bound matter will be assumed to be 
the virial velocity of the system. 
\begin{equation}
\label{eq:v_vir}
v_{vir} = \sqrt{\frac{GM_{Dyn}}{2 r_{1/2}}}.
\end{equation}
where $M_{Dyn}$ is the dynamical mass within a half-light radius.
Using the matter profiles (\ref{eq:NFW}) and (\ref{eq:plummer}) for the dark and visible 
components respectively and assuming a virial velocity given by (\ref{eq:v_vir}) 
the predicted flux can be written as, 
\begin{equation}
\label{eq:DsphFlux}
\frac{dN_{511}}{dt~dA} = \frac{\kappa}{d^2} v_{vir} 
\int \frac{\rho_0 \rho_s r_s r~dr}{\left(1 + (r/r_{1/2})^2\right)^{5/2} (1 + r/r_s)^2}
\end{equation}
Unlike WIMP annihilation models the 511 keV flux cannot be formulated purely 
in terms of the J-factor as it is also dependent on the visible matter profile. 
It is thus necessary for us to reformulate expression (\ref{eq:DsphFlux}) in terms of 
observable quantities. 

As the dwarf spheroids are dark matter dominated 
\cite{Wilkinson:2004fz, Strigari:2008ib, Irwin:1995} 
we will take the dynamical mass 
within a half-light radius to be primarily due to dark matter. We will also assume that the 
length scale associated with the dark matter halo is much greater than that associated with 
the visible matter (ie. $r_s >> r_{1/2}$). In this case we can integrate the NFW profile and 
express the dark matter scale density may be 
written as,
\begin{equation}
\rho_s r_s \approx \frac{M_{\rm Dyn}}{2\pi r_{1/2}^2}.
\end{equation}
The central density of the visible matter may be estimated by integrating 
the density profile in expression (\ref{eq:plummer}) and formulating the result in terms 
of the stellar mass within a half-light radius. This gives a central density of, 
\begin{equation}
\label{eq:vis_dens}
\rho_{0} \approx \frac{3 M_{\star}}{8\sqrt{2}\pi r_{1/2}^3}.
\end{equation} 
We may then represent expression (\ref{eq:DsphFlux}) as, 
\begin{eqnarray} 
\label{eq:FluxInt}
\frac{dN_{511}}{dt~dA} &=& 
\frac{3 \kappa}{16\sqrt{2}\pi^2 d^2}  \frac{v_{vir} M_{Dyn} M_{\star}}{r_{1/2}^5}  \\
&\times& \int \frac{ r~dr}{\left(1 + (r/r_{1/2})^2\right)^{5/2} (1 + r/r_s)^2}. \nonumber
\end{eqnarray}
If the dark matter distribution is sufficiently large relative to the visible matter 
that we may neglect the long range behaviour of the NFW profile then the integration 
of expression (\ref{eq:FluxInt}) is particularly simple, and it is given by
\begin{equation}
\label{eq:511flux}
\frac{dN_{511}}{dt~dA} \approx 
\frac{\kappa M_{Dyn} M_{\star}}{32 \pi^2 d^2 r_{1/2}^{3}} 
\sqrt{\frac{GM_{Dyn}}{r_{1/2}}}.
\end{equation}
So that we may estimate the total flux from a dwarf spheroid based on its 
observed dynamical mass, stellar mass and half light radius. The required 
parameters for the majority of galaxies considered in \cite{Siegert:2016ijv} may be 
found in \cite{McConnachie:2012}\footnote{This sample does not include the Reticulum II 
dwarf which shows a relatively strong 511 keV line as well as several others included in the 
analysis of \cite{Siegert:2016ijv}. It does however cover a wide range of the known 
dwarfs and there is no reason to expect that it is not a sufficiently representative 
subgroup of the known dwarfs.}. It is therefore possible to use expression 
(\ref{eq:511flux}) to predict a 511 keV line flux (normalized to that of the 
milky way) for each of the dwarf galaxies for which the relevant parameters 
are available. 

\begin{figure}
\includegraphics[width=\linewidth]{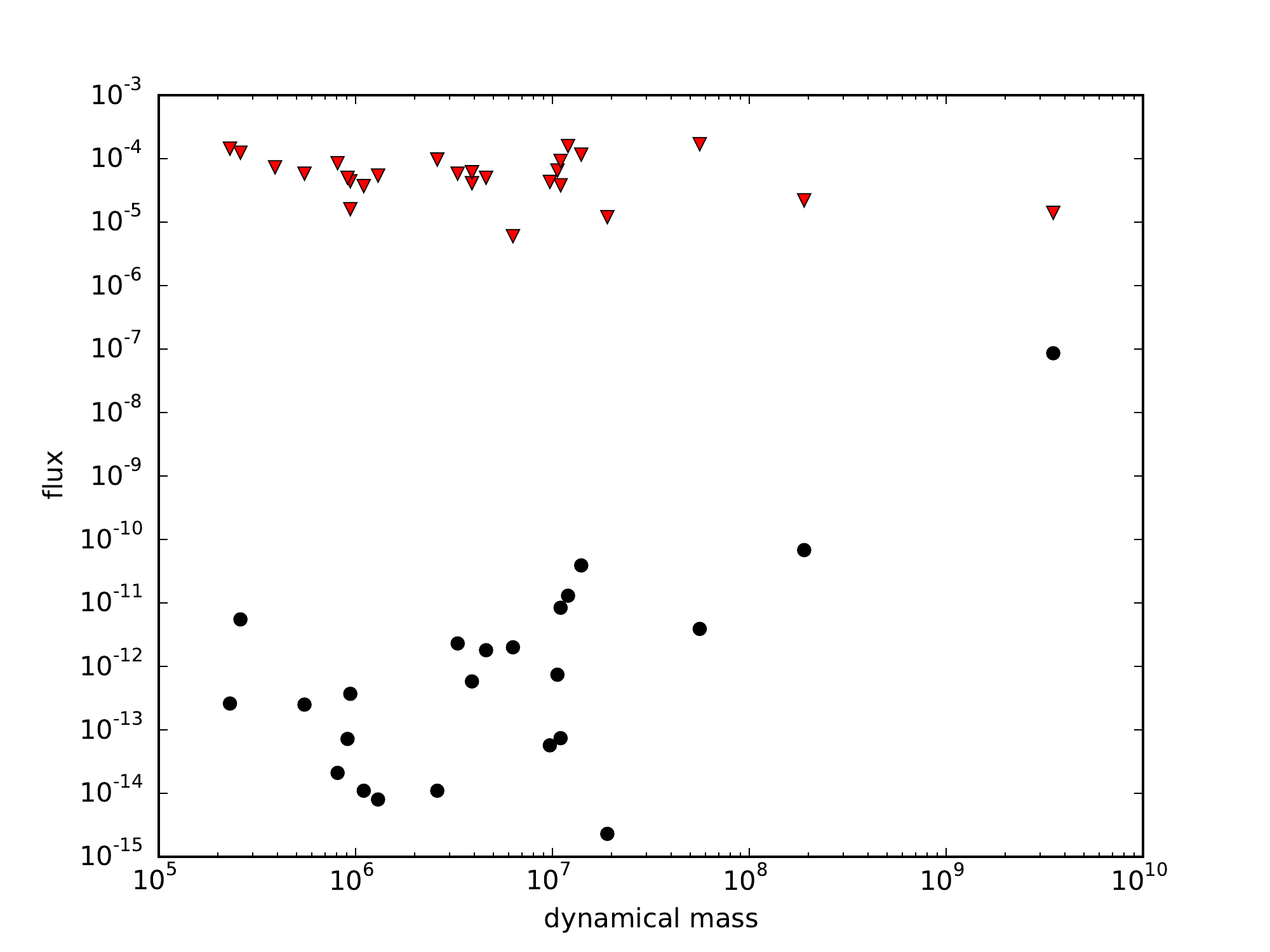}
\caption{Total 511 keV photon flux (in cm$^{-2}$s$^{-1}$ predicted from the dwarf 
spheroid galaxies described in \cite{McConnachie:2012} as a function of 
dynamical mass. The red triangles 
represent the observed flux or its upper limit in cases where no significant 511 
keV emission is detected. The black squares give the approximate 511 keV flux 
from a population of quark nuggets capable of providing the observed galactic 511 
keV derived from expression (\ref{eq:511flux}).}
\label{fig:flux_lim}
\end{figure}

Evaluating expression (\ref{eq:511flux}) for all the dwarf galaxies for which the 
relevant physical parameters are available in \cite{McConnachie:2012} results 
in the data shown in figure \ref{fig:flux_lim}.
As can easily be seen in the results presented in figure \ref{fig:flux_lim} the 511 keV 
flux generated by a population of quark nuggets capable of explaining the galactic 
511 keV line will be well below the observational limits. This is due to the fact that 
the interaction rate of the nuggets is set by the product of their density with the 
local visible matter density. In this case the predicted emissivity does not scale 
simple with the J-factor and, despite being dark matter dominated, the dwarf 
spheroid galaxies do not present a particularly attractive target for this type of 
dark matter model. 

\begin{figure}
\includegraphics[width=\linewidth]{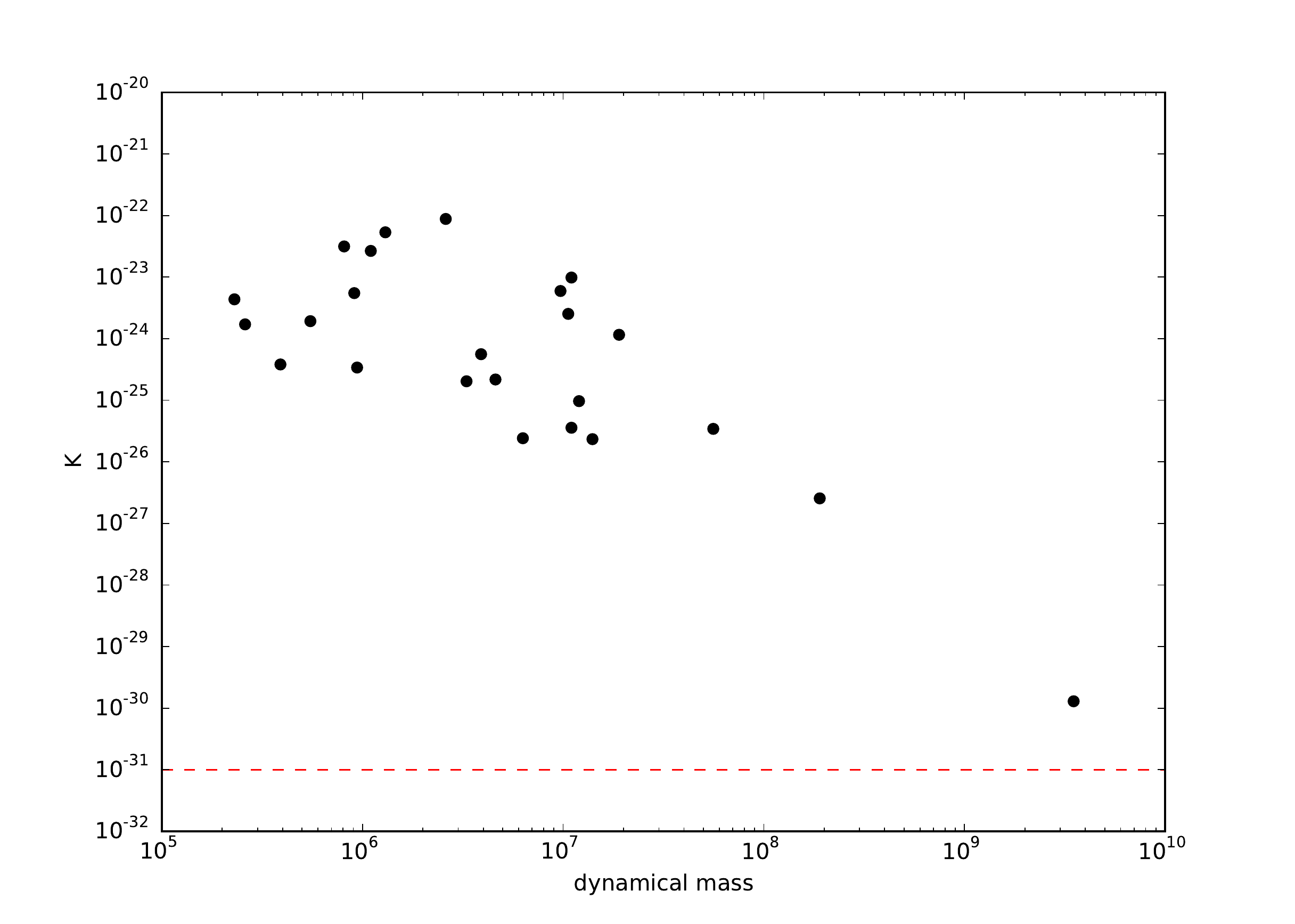}
\caption{Black dots show the values of the 511 keV line emission coefficient ($\kappa$) 
extracted from each of the dwarf spheroid galaxies considered in \cite{Siegert:2016ijv} 
and described in \cite{McConnachie:2012}. The value of $\kappa$ predicted if 
emission from electron-positron annihilation within antiquark nuggets saturates the 
galactic 511 keV line is shown by a red dashed line. }
\label{fig:exclusion}
\end{figure}

Adopting the flux limits cited in \cite{Siegert:2016ijv} and the physical 
parameters of \cite{McConnachie:2012} allows us to translate the 511 keV 
flux limits into constraints on the parameter $\kappa$ through equation 
(\ref{eq:511flux}). Figure \ref{fig:exclusion} shows the range of $\kappa$ values 
extracted from the dwarf spheroid galaxies as well as that obtained from the 
Milky Way. It is easily seen that the strength of the galactic 511 keV line 
produces a limit on the emission coefficient $\kappa$ at least an order of 
magnitude at least an order of magnitude below the best possible constraints 
obtained from the dwarf spheroids. 

Finally, it should be noted that, in this model, 511 keV  
emission from the dark matter dominated and gas poor dwarf galaxies is  mainly 
limited by the lack of visible matter with which the nuggets can interact. 
As such, the correlation between visible brightness and 511 keV line strength inferred 
in \cite{Siegert:2016ijv} does not contradict a possible dark matter origin. 
This is in contrast to models in 
which 511 keV emission is driven purely by dark sector physics in which case 
the arguments of \cite{Siegert:2016ijv} can be used to argue against 511 keV 
emission from dark matter.

\section{Conclusion}\label{sec:conclusion} 
The model we have advocated here was originally invented as a 
simple and natural explanation of the observed relation: 
$\Omega_{\rm dark} \simeq  \Omega_{\rm visible}$ 
by postulating that both elements originate from one and the same QCD epoch and proportional to one and the same $\Lambda_{\rm QCD}$ scale, see short review articles \cite{Lawson:2013bya} regarding the observational constraints  and \cite{Zhitnitsky:2016cir} regarding the formation stage during the QCD transition. 	
The immediate consequence of this proposal is the presence of antimatter 
in the form of macroscopically large antiquark nuggets. 
An equal portion of matter and antimatter in our universe does not contradict the 
conventional and naive arguments on the nearly total observational  absence of antimatter 
as has been explained in section \ref{sec:QNDM}. 

This model has a single fundamental parameter, the mean baryon number of a nugget  
$\langle B\rangle $,  which scales all observational consequences.  
If the value of $\langle B\rangle $ is sufficiently large the model is consistent with all 
known astrophysical, cosmological, satellite and ground based constraints as 
highlighted in section \ref{sec:QNDM}. 
Furthermore, in a number of cases the predictions of this model are very close to 
presently available limits, and very modest improvements on those constraints may lead to 
the discovery of the nuggets. Even more than that: there are a number of frequency bands 
where some excess of emission is observed, and this model may explain some portion, or 
even the entire excess of the observed radiation in these frequency bands.
 
In the present work we have explicitly demonstrated that this model is consistent 
with the non-detection of a 511 keV line from dwarf satellite galaxies. This should 
be contrasted with more conventional dark matter annihilation models for origin of the 
galactic 511 keV line 
(see the original paper \cite{Boehm:2003bt} and review article \cite{Prantzos:2010wi})
which the analysis of \cite{Siegert:2016ijv} shows to contradict the observed levels of 
511 keV emission from nearby dwarf spheroids. The crucial difference 
between our model and the large number of WIMP based models 
reviewed in \cite{Prantzos:2010wi}  
is that 511 keV emission strength is proportional to both the visible and 
dark matter densities as formula (\ref{flux1}) states.
In the case of 511 keV emission originating from the annihilation of 
light WIMP dark matter the positronium  
production rate is not sensitive to the visible matter density but depends 
exclusively on the dark matter J-factor according to eq. (\ref{J}). Precisely this fundamental difference in  
morphological properties between the models determines the drastic variation  
in the $e^+e^- $ annihilation  rate. 
  
One should also add that in our model the positrons fill the electrosphere of an antiquark 
nuggets with typical atomic velocities, such that the dominant mechanism of   
annihilation is through positronium formation in perfect agreement with observations.  This 
should be contrasted with the other dark matter explanations of the 511 keV line reviewed 
in \cite{Prantzos:2010wi} when the typical positron energies are order of MeV rather than 
atomic  (eV scale) energies. Furthermore, our model predicts  that  
the morphology of the 511 keV line should not be perfectly spherical as the 
rate depends on the visible density distribution according to (\ref{flux1}). These 
morphological features are also consistent with observations reviewed in  
\cite{Prantzos:2010wi}.
  
  \section*{Acknowledgments}
 
This work was supported in part by the National Science and Engineering
Research Council of Canada.


\begin{thebibliography}{99} 
\bibitem{Siegert:2016ijv} 
T.~Siegert, R.~Diehl, A.~C.~Vincent, F.~Guglielmetti, M.~G.~H.~Krause and C.~Boehm,
A\&A {\bf  595},  A25 (2016)
arXiv:1608.00393 [astro-ph.HE].

\bibitem{Prantzos:2010wi} 
  N.~Prantzos {\it et al.},
  Rev.\ Mod.\ Phys.\  {\bf 83}, 1001 (2011)
  [arXiv:1009.4620 [astro-ph.HE]].

\bibitem{Boehm:2003bt} 
  C.~Boehm, D.~Hooper, J.~Silk, M.~Casse and J.~Paul,
  Phys.\ Rev.\ Lett.\  {\bf 92}, 101301 (2004)
  [astro-ph/0309686].

\bibitem{Hooper:2003sh} 
  D.~Hooper, F.~Ferrer, C.~Boehm, J.~Silk, J.~Paul, N.~W.~Evans and M.~Casse,
  Phys.\ Rev.\ Lett.\  {\bf 93}, 161302 (2004)
  [astro-ph/0311150].
  
\bibitem{Zhitnitsky:2002qa} 
  A.~R.~Zhitnitsky,
  JCAP {\bf 0310}, 010 (2003)
  [hep-ph/0202161].
  
\bibitem{Oaknin:2003uv} 
  D.~H.~Oaknin and A.~Zhitnitsky,
  Phys.\ Rev.\ D {\bf 71}, 023519 (2005)
  [hep-ph/0309086].

\bibitem{Oaknin:2004mn} 
  D.~H.~Oaknin and A.~R.~Zhitnitsky,
  Phys.\ Rev.\ Lett.\  {\bf 94}, 101301 (2005)
  [hep-ph/0406146].

\bibitem{Zhitnitsky:2006tu} 
  A.~Zhitnitsky,
  Phys.\ Rev.\ D {\bf 76}, 103518 (2007)
  [astro-ph/0607361].
  
\bibitem{Lawson:2013bya} 
  K.~Lawson and A.~R.~Zhitnitsky,
  ``Quark (Anti) Nugget Dark Matter,''
  arXiv:1305.6318 [astro-ph.CO].

\bibitem{Liang:2016tqc} 
  X.~Liang and A.~Zhitnitsky,
  Phys.\ Rev.\ D {\bf 94}, 083502 (2016)
  [arXiv:1606.00435 [hep-ph]].

\bibitem{Zhitnitsky:2016cir} 
  A.~Zhitnitsky,
  ``Beyond WIMPs: the Quark (Anti) Nugget Dark Matter,''
  arXiv:1611.05042 [hep-ph].

  
\bibitem{Witten:1984}
E.~Witten
Phys.Rev. D {\bf{30}}, 272 (1984).

\bibitem{Zhitnitsky:2006vt} 
  A.~Zhitnitsky,
  Phys.\ Rev.\ D {\bf 74}, 043515 (2006)
  [astro-ph/0603064].

\bibitem{Forbes:2006ba} 
  M.~M.~Forbes and A.~R.~Zhitnitsky,
  JCAP {\bf 0801}, 023 (2008)
  [astro-ph/0611506].

\bibitem{Lawson:2007kp}
K.~Lawson and A.~R.~Zhitnitsky, 
JCAP {\bf{0801}}, 022 (2008),
[arXiv:0704.3064 [astro-ph]].

\bibitem{Forbes:2008uf} 
  M.~M.~Forbes and A.~R.~Zhitnitsky,
  Phys.\ Rev.\ D {\bf 78}, 083505 (2008)
  [arXiv:0802.3830 [astro-ph]].
    
\bibitem{Forbes:2009wg} 
M.~M.~Forbes, K.~Lawson and A.~R.~Zhitnitsky,
Phys.\ Rev.\ D {\bf 82}, 083510 (2010).
 [arXiv:0910.4541 [astro-ph.GA]].

\bibitem{Lawson:2012zu} 
  K.~Lawson and A.~R.~Zhitnitsky,
  Phys.\ Lett.\ B {\bf 724}, 17 (2013)
  [arXiv:1210.2400 [astro-ph.CO]].

\bibitem{McMillan:2011}
P.~J.~McMillan, 
MNRAS {\bf{414}}, 2446 (2011). 

\bibitem{Navarro:1995iw} 
J.~F.~Navarro, C.~S.~Frenk and S.~D.~M.~White,
Astrophys.\ J.\  {\bf 462}, 563 (1996)
[astro-ph/9508025]. 

\bibitem{Grcevich:2009gt} 
J.~Grcevich and M.~E.~Putman,
Astrophys.\ J.\  {\bf 696}, 385 (2009)
Erratum: [Astrophys.\ J.\  {\bf 721}, 922 (2010)]
[arXiv:0901.4975 [astro-ph.GA]].

\bibitem{Mashchenko:2005bj} 
  S.~Mashchenko, A.~Sills and H.~M.~P.~Couchman,
  Astrophys.\ J.\  {\bf 640}, 252 (2006).
  
\bibitem{Irwin:1995}
M.~Irwin and D.~Hatzidimitriou 
MNRAS {\bf{277}}, 1354 (1995). 

\bibitem{Wilkinson:2004fz} 
  M.~I.~Wilkinson, J.~T.~Kleyna, N.~W.~Evans, G.~F.~Gilmore, M.~J.~Irwin and E.~K.~Grebel,
  Astrophys.\ J.\  {\bf 611}, L21 (2004)
  [astro-ph/0406520].

\bibitem{Strigari:2008ib} 
L.~E.~Strigari, J.~S.~Bullock, M.~Kaplinghat, J.~D.~Simon, M.~Geha, B.~Willman 
and M.~G.~Walker,
Nature {\bf 454}, 1096 (2008)
[arXiv:0808.3772 [astro-ph]].
  
\bibitem{McConnachie:2012}
A.~W.~McConnachie, 
AJ {\bf{144}}, 4 (2012).


\end{thebibliography}
\end{document}